# A Freeform Dielectric Metasurface Modeling Approach Based on Deep Neural Networks

*Sensong An[1], Bowen Zheng[1], Mikhail Y. Shalaginov[2], Hong Tang[1], Hang Li[1], Li Zhou[1], Jun Ding[3], Anuradha Murthy Agarwal[2], Clara Rivero-Baleine[4], Myungkoo Kang[5], Kathleen A. Richardson[5], Tian Gu[2], Juejun Hu[2], Clayton Fowler[1,\*] and Hualiang Zhang[1,\*]*

[1]Department of Electrical & Computer Engineering, University of Massachusetts Lowell, Lowell, Massachusetts 01854, USA

[2]Department of Materials Science & Engineering, Massachusetts Institute of Technology, Cambridge, Massachusetts 02139, USA

[3]Shanghai Key Laboratory of Multidimensional Information Processing, East China Normal University, Shanghai 200062, China

[4]Lockheed Martin Corporation, Orlando, Florida 32819, USA

[5]CREOL, University of Central Florida, Orlando, Florida 32816, USA

*clayton_fowler@uml.edu   *hualiang_zhang@uml.edu

ABSTRACT: Metasurfaces have shown promising potentials in shaping optical wavefronts while remaining compact compared to bulky geometric optics devices. Design of meta-atoms, the fundamental building blocks of metasurfaces, relies on trial-and-error method to achieve target electromagnetic responses. This process includes the characterization of an enormous amount of different meta-atom designs with different physical and geometric parameters, which normally demands huge computational resources. In this paper, a deep learning-based metasurface/meta-atom modeling approach is introduced to significantly reduce the characterization time while maintaining accuracy. Based on a convolutional neural network (CNN) structure, the proposed deep learning network is able to model meta-atoms with free-form 2D patterns and different lattice sizes, material refractive indexes and thicknesses. Moreover, the presented approach features the capability to predict meta-atoms' wide spectrum responses in the timescale of milliseconds, which makes it attractive for applications such as fast meta-atom/metasurface on-demand design and optimization.

Metasurfaces, the 2D version of metamaterials, provide a novel platform for the realization of ultra-thin and large-scale optical components and systems. By manipulating the geometry of individual meta-atoms, pre-assigned responses (e.g. phase and amplitude) can be realized at the unit-cell level over the flat surface for full control of light propagation and the design of meta-devices. The key challenge in the field of metasurface/meta-atom design is the non-intuitive design process, which makes it difficult to find optimal design parameters to meet specific requirements. Current design approaches include trial-and-error methods and inverse design methods based on optimization algorithms or deep neural networks (DNNs). For the traditional trial-and-error method, a commonly adopted design process includes a complete exploration of all design spaces and careful selection of results that fit the design requirements. In this case, the design time is determined by simulation time of each single design and the number of design degrees of freedom. Therefore, a full exploration of all design parameters is unrealistic when massive design degrees of freedom must be taken into consideration (e.g. free-form meta-atom designs) while a fast evaluation tool is not available. As for the inverse design approaches, either based on optimization algorithms [1-4] or neural networks [5-13], their converging speed and design accuracy largely rely on the simulation speed and accuracy of the local or commercial solver that is cascaded to the optimizer. One exception is the very recently emerged deep learning metasurface design approach based on generative adversarial networks (GANs) [14-18], which incorporates a self-evolving Critic, rather than a well-trained DNN simulator to evaluate the performance of the generated designs. However, since a GAN requires a noise prior as part of the input, the corresponding output designs have unstable performance and thus still need to be verified by a simulator. Therefore, modeling and characterization tools play a pivotal role in almost all current metasurface/meta-atom design approaches, reliable and time efficient modeling tools are always in need and are being heavily investigated in the metasurface design field to meet the stringent designed requirements imposed by next generation meta-devices featuring free-form shapes and multi-functionalities.

One approach is to develop analytical effective medium models, such as the Lewin model [19] and the GEM model [20]. Although these models are simple and efficient, they only tackle metamaterials with the shape of microspheres under the long-wavelength approximations. Another widely adopted approach relies on iterative numerical full-wave simulations based on different methods including FEM (finite-element method), FDTD (finite-difference time-domain) and FIT (finite integration technique). This approach provides accurate results but requires considerable computing resources. Different from analytical models (with a lot of limitations) and full-wave simulations (universal but time-consuming), a data-driven modeling tool based on deep neural networks (DNNs) [9, 11, 13, 21] emerged recently and has been proven to be accurate and timely-efficient compared to conventional approaches. Previous works have employed fully-connected layers (FCLs) to realize the accurate spectrum response predictions on nanophotonic structures with bulk layers[6], cavities[22] and meta-atoms in the shapes of cylinders [8, 9, 12, 13], elliptic cylinders [23], spheres [11] and bars [5, 10] constructed with plasmonic [5, 9, 10] or all-dielectric [6, 8, 12, 13] materials. While most works [5, 6, 8-11, 23] have been focused on amplitude response predictions, some very recent works[13] have demonstrated that the neural networks are also capable of predicting meta-atoms' phase responses, which is crucial considering most optical applications required full manipulation of incident light. After being fully-trained with sufficient data, the DNN models are highly accurate and able to generate EM responses on the time scale of milliseconds, which enables fast on-demand meta-atom/metasurface designs. However, there are some main issues with these existing networks constructed with FCLs. Firstly, these works mainly deal with simple meta-atom structures that can be easily described by 3-5 parameters, which has limited meta-atoms' capabilities in achieving high efficiency and broad phase coverage and the application of compounded metasurfaces. Secondly, these DNN models are operating in a very restricted design space. Design parameters including lattice sizes, meta-atom thicknesses and material properties are fixed in these networks. Once one of these design parameters was changed, the data needs to be re-collected and the model has to be re-trained, which can be time-consuming. This weak generalization ability also limits the efficacy of the current DNN approach.

In this paper, we present a new DNN approach for the modeling and characterization of three-dimensional (3D) meta-atoms, which addresses both issues discussed above. To expend the design space and demonstrate the network's generality, our approach takes into consideration almost all design spaces of a meta-atom, which includes the meta-atom's two-dimensional (2D) geometrical pattern, material index, thickness and lattice size. After trained with sufficient data, the proposed network is able to generate accurate phase and amplitude predictions of meta-atoms with complex shapes across a wide spectrum. Furthermore, we demonstrated our network's generalization abilities by testing it with meta-atoms which contain features that never existed in the training data. To show the efficacy of the proposed method, it has been applied for practical metasurface/meta-device design and optimization. The performance of resulting metasurface/meta-device prototypes corroborates that the presented network achieved two important features for DNN-based meta-atom modeling: 1) fast and accurate performance evaluation of topologically complex meta-atoms and metasurfaces; and 2) a modeling tool that covers the full design space of 3D meta-atoms. It is envisioned that the proposed deep learning network can be readily applied to various meta-atom modeling and optimization tasks, as well as being extended to other fields such as the characterization and design of dielectric resonator antennas, optical circuits and chiral metamaterials.

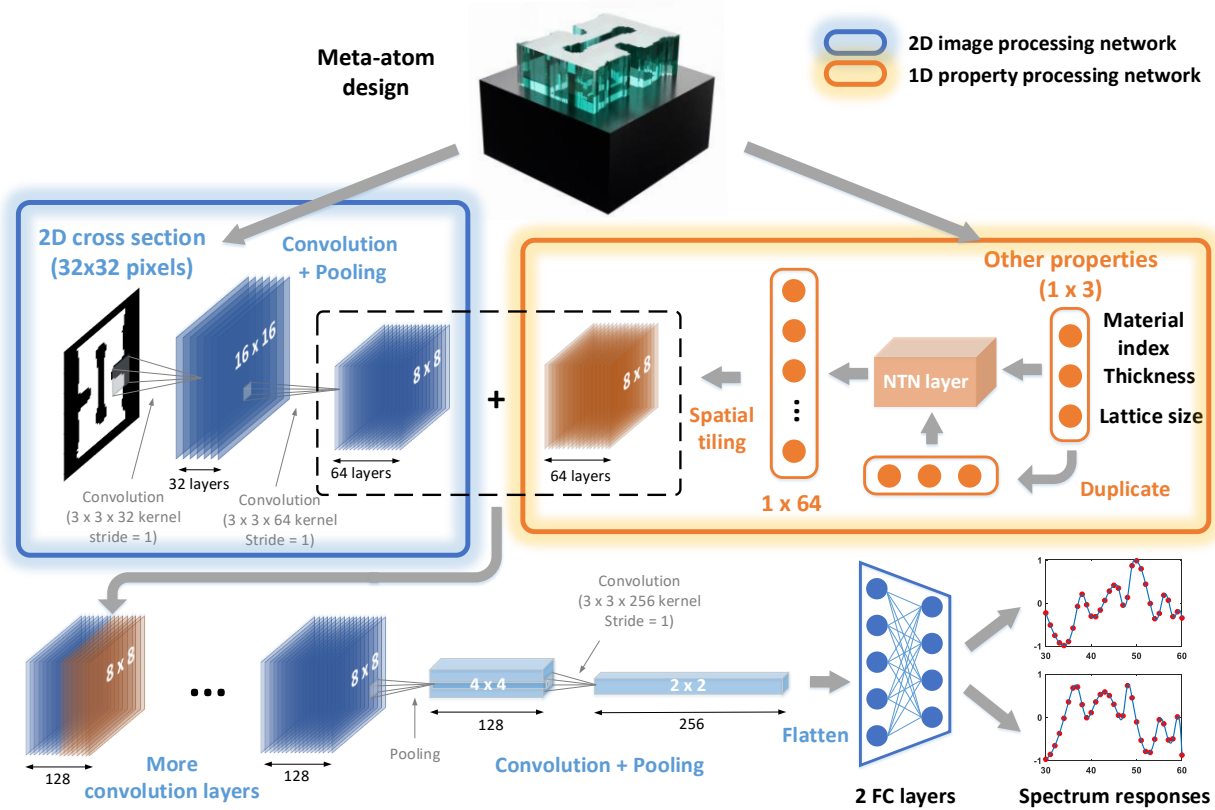

**Figure 1. Network architecture.** Meta-atoms design parameters were split into 1D property parameters and 2D cross sectional images and processed through a 1D property processing network (circled in yellow) and a 2D image processing network (circled in blue). 2D image (32 x 32 pixels) were processed with 2 convolution layers and then combined with 1D properties (with the size of 8 x 8 x 64 through spatial tiling). Combined results (8 x 8 x 128) are further processed with more convolution and pooling layers before finally flatten into a 1D array (1 x 1024). After being processed with 2 more FCLs, the real/imaginary part (1x51) of the transmission coefficient over the spectrum of 30-60THz were ready for evaluation. All convolution layers in the network are followed by a batch normalization layer. More detailed network architecture can be found in section I of the supporting information.

**Network Architecture.** To address the two goals discussed above, a predicting neural network (hereinafter called the 'PNN') was constructed based on a CNN architecture (Figure. 1). The PNN aims to uncover the hidden relationship between meta-atom models and their spectrum responses and thus predict accurate responses for given met-atom designs. The meta-atom model under evaluation consists of a freeform dielectric structure (preferably with a higher refractive index) sitting on a square-shaped dielectric substrate (preferably with a lower refractive index). To prove the proposed method is not limited to meta-atoms with certain shapes or materials, we parameterized the meta-atom structure's 2D pattern and its other properties including lattice size, thickness and refractive index. All these properties were combined and designated as the input of the PNN. To reconcile the huge dimension mismatch between the 2D cross-section (in this case is an image composed of 32x32 pixels) and other properties (contain 3 elements), the combined input was processed using a 2D image processing network (circled in blue in Fig. 1) and a 1D property processing network (circled in yellow in Fig. 1), respectively. For the 1D property processing network, to speed up the training process and achieve better accuracy, we applied a Neural Tensor Network (NTN) layer [10, 13, 24] to deal with the relational information provided with the 1D input properties. The output of the NTN layer was then replicated over the spatial dimensions [25] of the output given by the 2D image processing network. Two outputs were then concatenated together and processed with more convolution and pooling layers. After flattening the output of the CNNs and pass it through two fully connected layers, the predicted real and imaginary parts of the transmission coefficient were generated and ready for evaluation. Without loss of generality, the spectra of interest were set to be from 30 to 60 THz (5 μm to 10 μm in wavelength).

Over 50,000 groups of quasi-freeform meta-atom patterns were randomly generated using the "needle drop" approach (Supporting information Section II), while the other parameters are created randomly within the ranges (all lengths in microns): thickness $\in [0.5, 1]$, refractive index $\in [3.5, 5]$, lattice size $\in [2.5, 3]$, since these ranges include ample samples of phase and amplitude coverage. The electromagnetic responses of these meta-atom models were then calculated in a FEM-based simulation tool and assigned as labels. Among these data, 70% are used during the training process, while the remaining 30% are used to evaluate the well-trained network. The spectrum response predictions generated with the PNN are compared with the labels to extract the error, which will be minimized during the training process (learning curves and Hyperparameters are included in supporting information Section III). When the training is completed, the average mean square error (MSE) for the real and imaginary part of the transmission coefficient is 0.00035 and 0.00023, respectively (equivalent to an average prediction deviation of 0.005 (amplitude) and 0.78 degrees (phase) at each single frequency point). An ablation analysis (Supporting information Section IV) is also carried out to justify the necessity of the different data processing approaches adopted in this network, including the use of NTN layers, batch normalization layers, spatial tiling and split real/imaginary component prediction method. It is concluded from the ablation analysis that removal of these layers resulted in either slower converging speed or lower final accuracy leading to inaccurate prediction results (Supporting information section III).

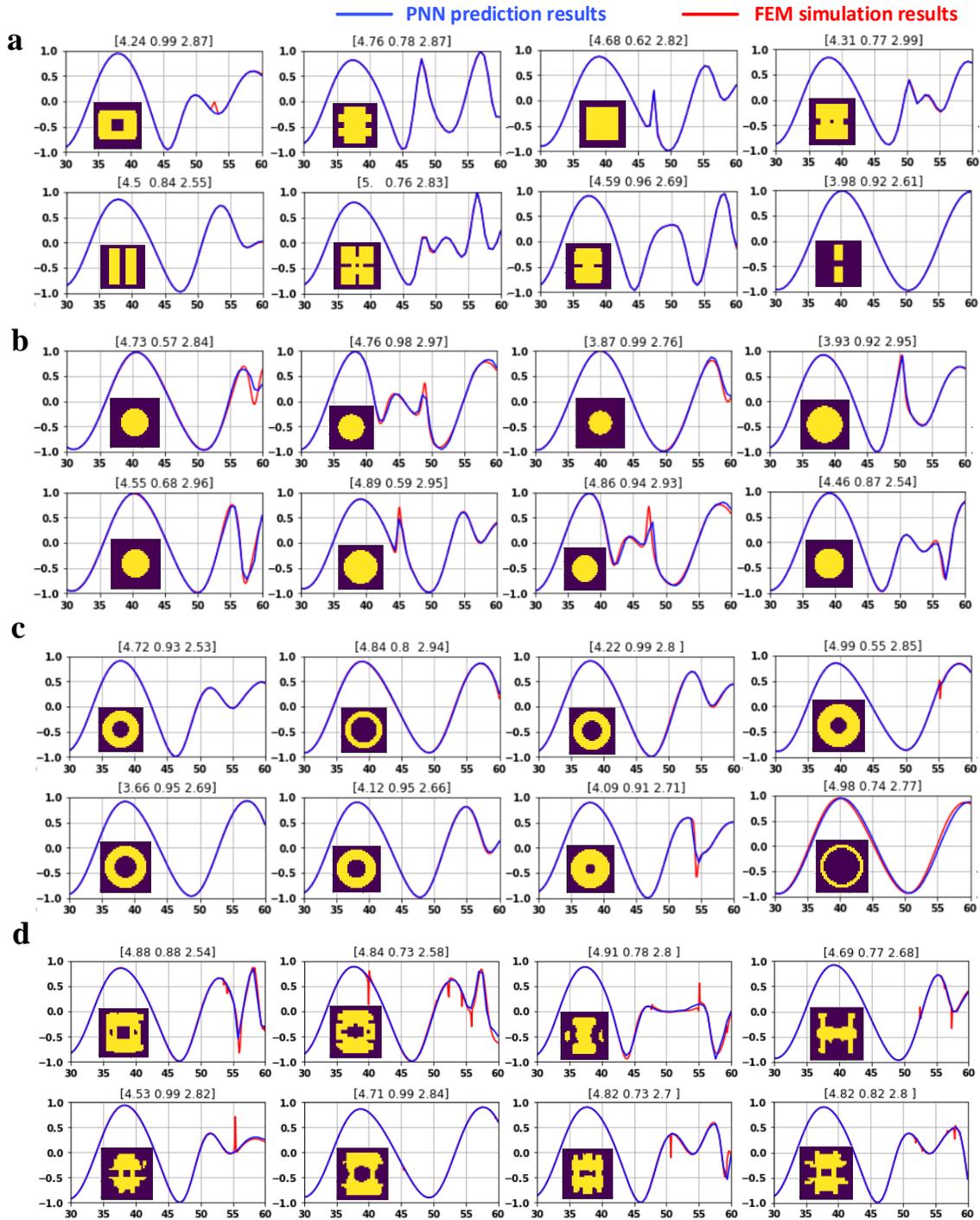

**Figure 2. PNN predictions compared to accurate results. a** PNN predictions on meta-atoms selected from the test dataset. **b** PNN predictions on circle-shaped meta-atoms. **c** PNN predictions on ring-shaped meta-atoms. **d** PNN predictions on slightly asymmetric meta-atoms. Blue curves represent the PNN predictions, while red curves are simulation results. Parameters including refractive index, meta-atom thickness and lattice size are shown on top of each subplot (in that order and lengths in μm). 2D cross sections of each meta-atom are included as insets. Only the real parts of the complex transmission coefficients are plotted for demonstrating purpose. Additional PNN prediction results can be found in supporting information section V.

**Results.** After the error is stabilized and the training is over, we employed the well-trained PNN to evaluate some randomly-selected meta-atom structures to visualize its prediction accuracy. Eight meta-atom samples (Fig. 2a) were randomly selected from the test dataset, the real parts of their complex transmission coefficients were evaluated (blue curves), and compared them with the results derived from FEM-based simulations tools (red curves). As indicated by the minimal training loss, the PNN prediction results agreed well with the full-wave electromagnetic simulations. Moreover, to verify generalization performance of the invented PNN and its adaptive ability to new, previously unseen data, we tested the PNN with some newly-generated meta-atom designs possessing features that did not exist in either training dataset or test dataset. This included meta-atoms with the shape of rings (Fig. 2b), circles (Fig. 2c) and even slightly asymmetric patterns (Fig. 2d). The radius defining the circles and rings in Fig. 2b and 2c are randomly generated, while the slightly asymmetric patterns in Fig. 2d are derived from a meta-atom design networks based on GAN [18]. Other parameters including refractive index, meta-atom thickness and lattice size are randomly selected within the preset range. Similar to Fig. 2a, the real parts of transmission coefficients of all designs in Fig. 2b-d are evaluated using the PNN (blue curves), and then compared with the full-wave electromagnetic simulation results (red curves). The PNN maintained its accuracy even with those meta-atoms that have previously unseen features, indicating its broad generalization ability. More importantly, once trained with enough data, the proposed PNN is able to accomplish the predictions on milliseconds, which makes it appealing for applications that are used to be prohibitively time-consuming using conventional methods (e.g. full-wave simulations). These practical applications (such as meta-device optimizations and meta-atom design platform evaluations) will be further discussed in the following section.

**Discussion.** Dielectric metasurface/meta-atom design platforms built with various construction materials[26-28] have featured lots of design degrees of freedom, including meta-atom's 2D patterns, refractive indexes, thicknesses and lattice sizes. Since most metasurfaces/meta-devices are composed of elements with the same lattice size, thickness and constructing material, choice of this parameter combination determines the possible overall phase and amplitude coverage. With inappropriate choices of these parameters, it is difficult, if not impossible to realize large phase coverages, even with a high degree of freedom for choosing meta-atom shapes. Meanwhile, most high-efficiency meta-devices, including lenses and beam deflectors require meta-atoms that can achieve full 2π phase coverage while maintaining high transmission efficiency. As a result, designers used to explore numerous parametric spaces for optimal parameter combinations that provides maximum phase and amplitude coverage. This design space exploration process is time consuming and inefficient (many times it is even impossible due to the prohibitively long process).

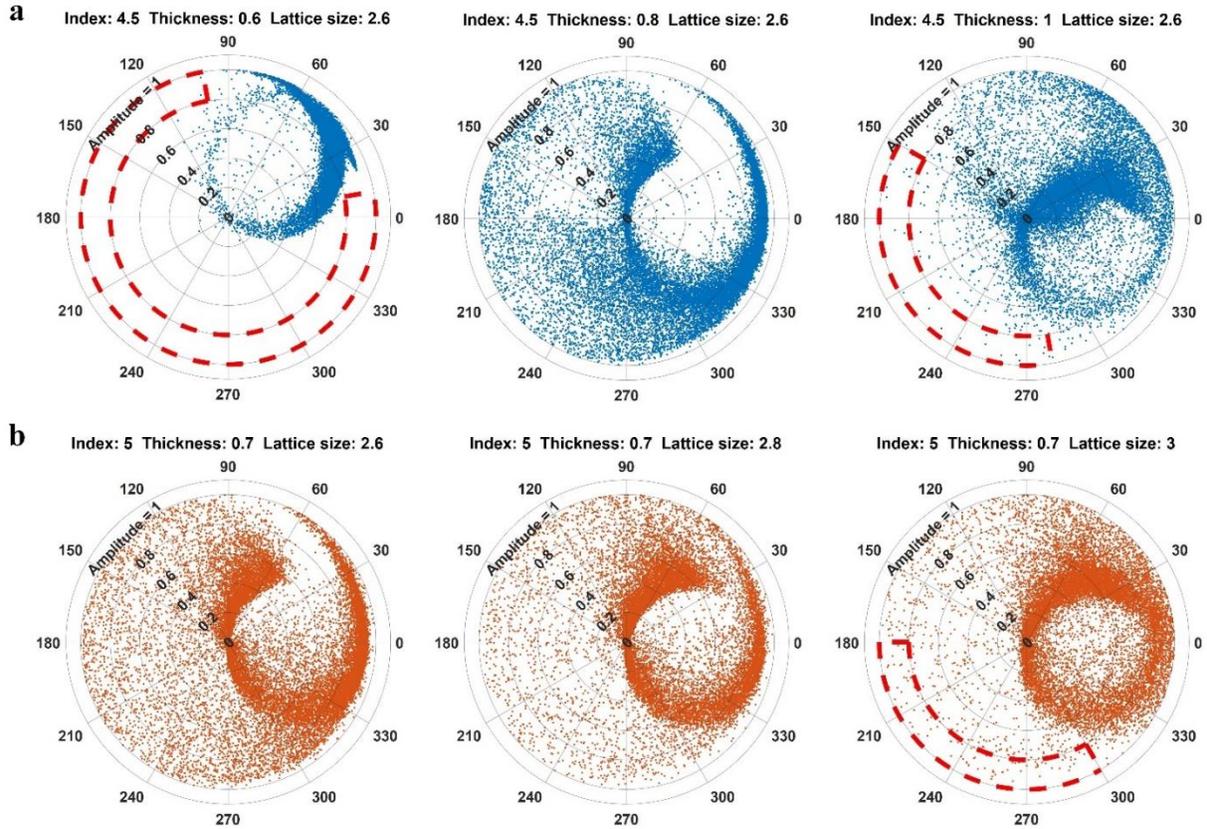

**Figure 3. Meta-atoms' EM performance evaluated using PNN. a** Phase and amplitude coverage with fixed index, lattice size and changing thicknesses. **b** Phase and amplitude coverage with fixed index, thickness and changing lattice sizes. Areas that are sparsely populated by high-efficiency candidates are circled in red dotted lines. Working frequency is set to be 57 THz in all six cases.

Alternatively, the PNN is able to evaluate the phase and amplitude responses of a given meta-atom parameter combination in a short, one-time calculation process, leading to unprecedented capability for meta-atom design. For example, meta-atoms with larger volume and refractive index can support more electromagnetic resonances, and are thus more likely to achieve high efficiency with full 2π phase coverage [29]. However, in certain circumstances increasing thickness or lattice sizes can lead to mismatch of meta-atom's intrinsic electric dipoles and magnetic dipoles and reduce the overall phase coverage. To address this issue, we randomly generate over 20,000 different meta-atom patterns and combine them with different refractive indexes, thicknesses and lattices sizes, and evaluate their performance at 57THz (i.e. 5.26μm in wavelength) using the proposed PNN. As shown in Fig. 3a, we fix the index to be 4.5 and lattice size to be 2.6μm, and evaluate the thickness's influence on the final phase coverage of the same group of meta-atom patterns. It is observed that the phase coverage increased at first, but started to descend when the thickness continued to increase. When the thickness is equal to 1μm, it's hard to pick a group of high efficiency meta-atoms in the area circled in red dashed line. Similarly in Fig. 3b, when fixing the index to be 5 and thickness to be 0.7μm, the meta-atom phase coverage drops as the lattice size increases from 2.6μm to 3μm. In both cases, the proper meta-atom design parameter combinations with maximum phase coverage can be identified in seconds using the presented PNN, which highlighted its efficacy in searching for new meta-atom designs.

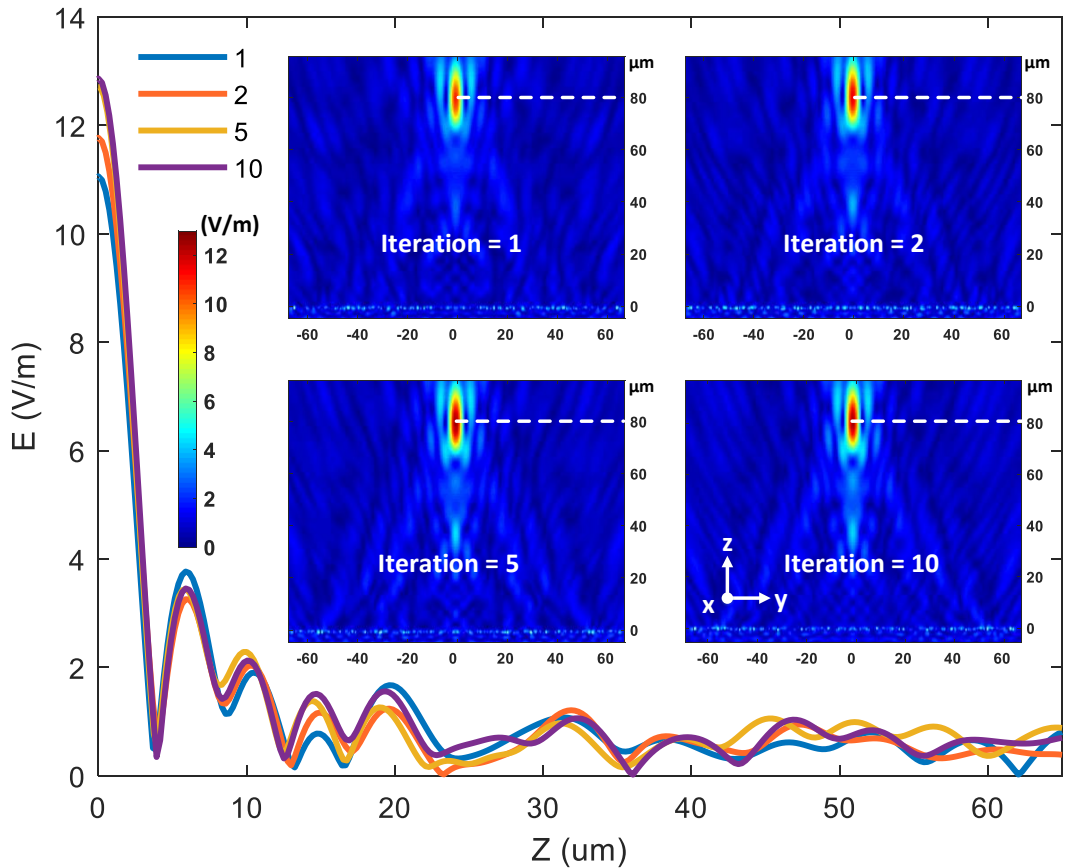

**Figure 4. Metalens optimization using the well-trained PNN.** Four different metalenses are derived using the GAN-PNN optimization network and then evaluated with full-wave simulations. The inset 2D images are E-field on the plane perpendicular to the metasurface. 1D curves are the E-field along white dotted lines in each figure. Full-wave simulation results show peak E-field increased with more optimization steps. Full-wave simulation results are derived using the time domain solver in FEM simulation tool CST.

Another application of the invented PNN is adopting it as an optimization tool for DNN-based meta-device designs. One major advantage of the DNN-based metasurface/meta-atom design approaches is that they are able to generate multiple designs with almost no cost. But all the generated designs need to be characterized and evaluated to identify the best fit design, which can be much more time-consuming than the design-generation process (depending on the applied simulation tool). The proposed PNN can greatly alleviate such an issue due to its fast response prediction capability. As a demonstration, we employed a well-trained GAN model[18] to design a transmissive meta-lens working at 57THz. The lens was composed of 50 by 50 meta-atoms with the lattice size of 2.6μm, which is equivalent to a full dimension of 130μm by 130μm. Focal length of this lens was set to be 80um, with an NA of 0.63. The lens was composed of dielectric meta-atoms (with the refractive index of 4.7 and thickness of 0.75μm) sitting on a dielectric substrate (with the refractive index of 1.4). After the phase mask of this lens was calculated, we trained a generative meta-atom design network [18] to generate meta-atom design for each unit cell within the metalens. The EM responses of the generated meta-atoms were then evaluated using the proposed PNN. This cascaded design-evaluate process was executed for several iterations so that the best meta-atom designs generated during these iterations were selected to assemble the final device. To verify the prediction accuracy of the PNN and the efficacy of this design-evaluate process, we employ this cascaded network (GAN + PNN) to run 1, 2, 5 and 10 iterations to generate 4 different meta-lenses (shown in supporting information section VI), and then tested their performance using full-wave simulation tool. The results are plotted in Fig. 4. The metalenses are placed in the x-y plane, with the optical axis along the z-axis. The simulated 2D electric fields of four different metalenses in the x-z plane, along with the 1D electric field along the x axis in the focal plane, are plotted in Fig. 4. The peak electric field amplitude at the center of the focal spot was improved with the increase of optimization iterations, validating that meta-atom designs with better performance (higher transmission and precise phase shift) have been identified with the help of PNN during the optimization iterations. Importantly, with this data driven approach, time taken for the evaluation process is largely reduced and comparable with the design generation time, which has solved the verification overhead problem suffered by the existing DNN-based approaches. A detailed time efficiency contrast analysis is also included in the supporting information section VII.

**Conclusion.** In this paper, a deep learning-based dielectric meta-atom modeling approach is proposed and demonstrated. Accurate spectrum responses of dielectric meta-atoms were derived using a novel CNN-based network structure. Compared with previous works, our approach has largely expanded the input degrees of freedom and has taken multiple parameters into account, including the shapes, thickness, lattice size and refractive index of the meta-atoms. It is validated that the proposed deep learning network processes strong generalization ability and is able to handle meta-atoms with features that doesn't exist in the training dataset. We have further demonstrated that the presented network can be adopted as an efficient modeling tool in various application scenarios, including rapid meta-atom design and meta-device optimization. It is envisioned that the proposed DNN-based methodology can be readily applied to other physics domains where a fast, accurate modeling tool is highly desired to provide the link between a broad and sophisticated parametric space and the corresponding physical responses.

# A Freeform Dielectric Metasurfaces Modeling Approach Based on Deep Neural Networks

*Sensong An[1], Bowen Zheng[1], Mikhail Y. Shalaginov[2], Hong Tang[1], Hang Li[1], Li Zhou[1], Jun Ding[3], Anuradha Murthy Agarwal[2], Clara Rivero-Baleine[4], Myungkoo Kang[5], Kathleen A. Richardson[5], Tian Gu[2], Juejun Hu[2], Clayton Fowler[1,\*] and Hualiang Zhang[1,\*]*

[1]Department of Electrical & Computer Engineering, University of Massachusetts Lowell, Lowell, Massachusetts 01854, USA

[2]Department of Materials Science & Engineering, Massachusetts Institute of Technology, Cambridge, Massachusetts 02139, USA

[3]Shanghai Key Laboratory of Multidimensional Information Processing, East China Normal University, Shanghai 200062, China

[4]Lockheed Martin Corporation, Orlando, Florida 32819, USA

[5]CREOL, University of Central Florida, Orlando, Florida 32816, USA

*clayton_fowler@uml.edu   *hualiang_zhang@uml.edu

In this Supporting Information, we provide further details on network modeling, network training and additional examples showcasing the performances of different networks. This Supplementary Information consists of the following Sections:

I. Detailed information of the network architecture

II. Data collection process

III. Hyperparameters used in the DNN training process

IV. Ablation analysis

V. Additional samples of the PNN

VI. Details of the metalens optimization process

VII. Computation efficiency analysis

## Section I – Detailed information of the network architecture

Fig. S1 illustrates the network architecture of the proposed PNN, with the data flow details included. The input meta-atom design was decomposed into a 2D image tensor (32 x 32) and a 1D property tensor (1 x 3), and then processed through two distinct networks for further feature extraction. The output of the Neural Tensor Network (NTN) constructed in the 1D property processing network is given by:

$$Output = f(e^T W^{[1:k]} e + b) \qquad (S1)$$

Where W (64 x 3 x 3) is the weight and b (1 x 3) is the bias. Outputs of these two networks are stacked (with the dimension of 8 x 8 x 128) and fed into the following 6 layers of CNNs and 2 layers of FCLs to be further processed. A batch normalization layer and a ReLU activation function is applied to the output tensor of each layer except for the last one. The real and imaginary parts of the complex transmission coefficient are predicted using two independent networks, which share the same network architecture. Dimensions of the convolutional kernels, fully-connected layers and pooling layers are included in Fig. S1. After the PNNs for the real & imaginary parts are constructed and well-trained, values of the weight and bias arrays in each network are fixed and saved for further use.

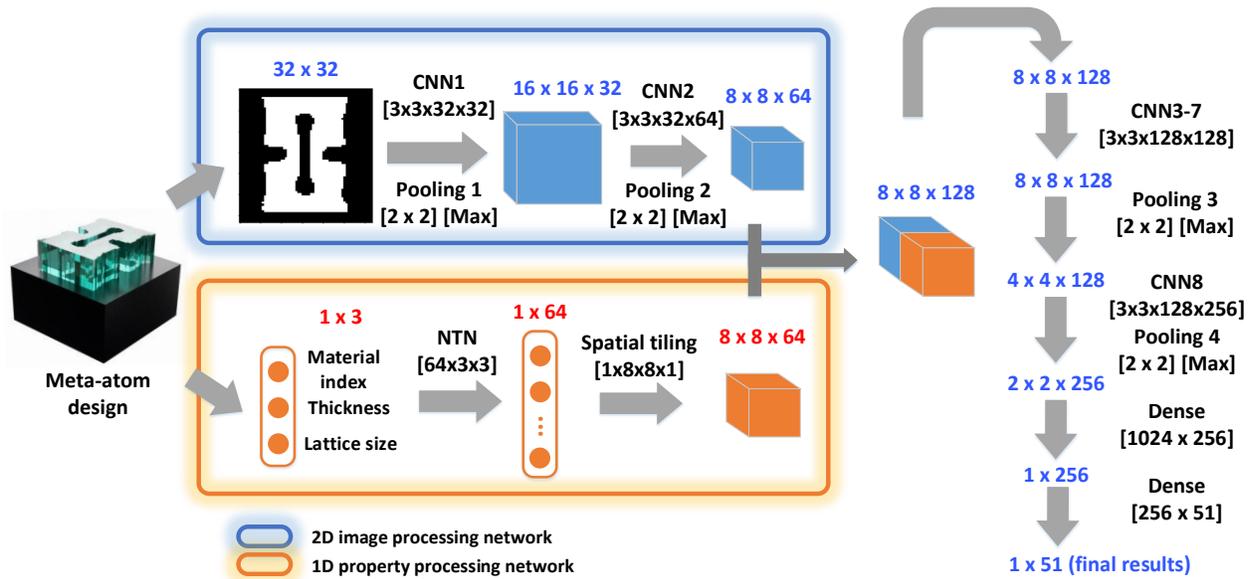

**Figure S1. The network architectures of the PNN.** Details of the eight convolutional layers, two fully-connected layers and one bilinear tensor layer are given in the figure. **b** Detailed network structures for the training of the meta-filter design network. The 1D property processing network and 2D image processing network are marked in different colors.

## Section II – Data collection process

As shown in Fig. S2a, the all-dielectric meta-atom consists of a dielectric component (preferably with a high refractive index, $n_1$) with the thickness of $t_1$ sitting on a dielectric substrate (preferably with a low refractive index $n_2$, in this case $n_2$ = 1.4) with a unit cell size of $l_1 \times l_1$ μm². The 2D pattern of each meta-atom was generated with the "Needle Drop" approach. Several (3 to 7) rectangular bars, with a minimum generative resolution of 1 pixel, were randomly generated and placed together within a square canvas (32 x 32 pixels) to form random patterns (Fig. S2b). To minimize inter-cell coupling, a minimum spacing of 4 pixels was applied between adjacent meta-atoms. To speed up the data-collection process, the all-dielectric components are only generated in the top left quadrant of each unit cell and then symmetrically replicated along *x* and *y* axes to form the whole pattern.

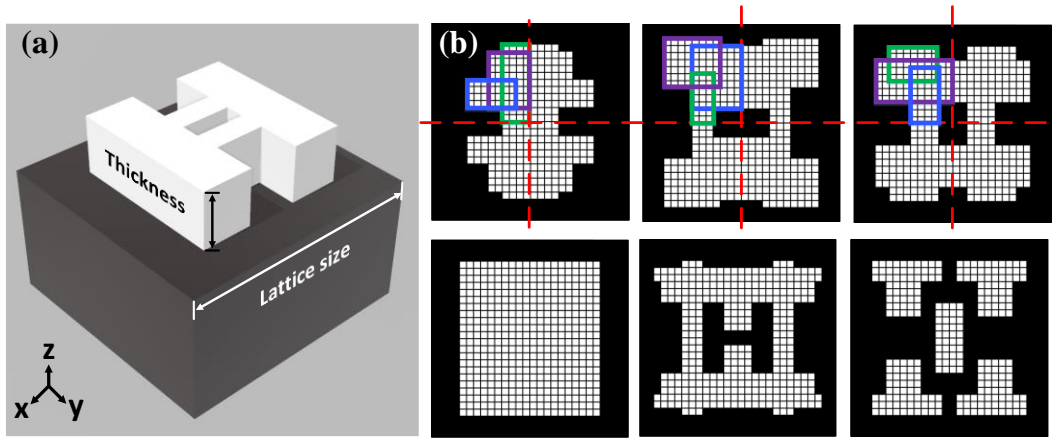

**Figure S2. Training data collection process. a** View of a randomly generated meta-atom. White represents high-index dielectric components, while black represents the low-index substrate. **b** Several examples of generated patterns. 2D patterns in x-y plane are meshed for better view, each mesh has a dimension of 1 by 1 pixel. Rectangles outlined in different colors represent distinct high-index "Needles" that were randomly generated and dropped on the top-left quadrant of the substrate canvas. Full patterns were completed by mirroring along the *x* and *y* axes.

The full-wave electromagnetic simulations were performed using a commercial FEM simulation tool, CST STUDIO SUITE. For each meta-atom, unit cell boundary conditions were employed to calculate the transmission and phase shift of a square lattice structure. Open boundaries are applied along both the negative and positive *z* directions, while an *x*-polarized plane wave was illuminated from the substrate side for each meta-atom. A total number of 53,000 meta-atoms with different shapes and refractive indexes, thicknesses, and lattice sizes were generated and simulated to find their wide-spectrum (30-60 THz) phase and amplitude responses. These meta-atoms are then labeled with their responses and documented for further training.

## Section III – Hyperparameters used in the DNN training process

Hyperparameters used in the training for both PNNs are shown in Table S1. The hardware consists of a quad-core CPU with 3.5 GHz clock speed, 64 Gigabytes of RAM and two NVidia 1080Ti GPUs. After 10,000 iterations, the average test set error stabilized at 0.0027 and 0.0025 for the real and imaginary part prediction networks, respectively (Table S1). With the current hardware setup, the training takes 48 hours for both PNNs before their error rates stabilize.

**Table S1. Hyperparameters used in the training of PNNs**

| Hyperparameters | PNN (real part) | PNN (Imaginary part) |
| --- | --- | --- |
| Training set size | 37100 | 37100 |
| Test set size | 15900 | 15900 |
| Optimizer | Adam | Adam |
| Learning rate | $10^{-4}$ | $10^{-4}$ |
| Batch size | 64 | 64 |
| Batch Norm. | Yes | Yes |
| Nonlinear activations | ReLU | ReLU |
| Iterations | 10000 | 10000 |
| Time taken | 48 h | 48 h |
| Error (train) | 0.0014 | 0.0012 |
| Error (test) | 0.0027 | 0.0025 |

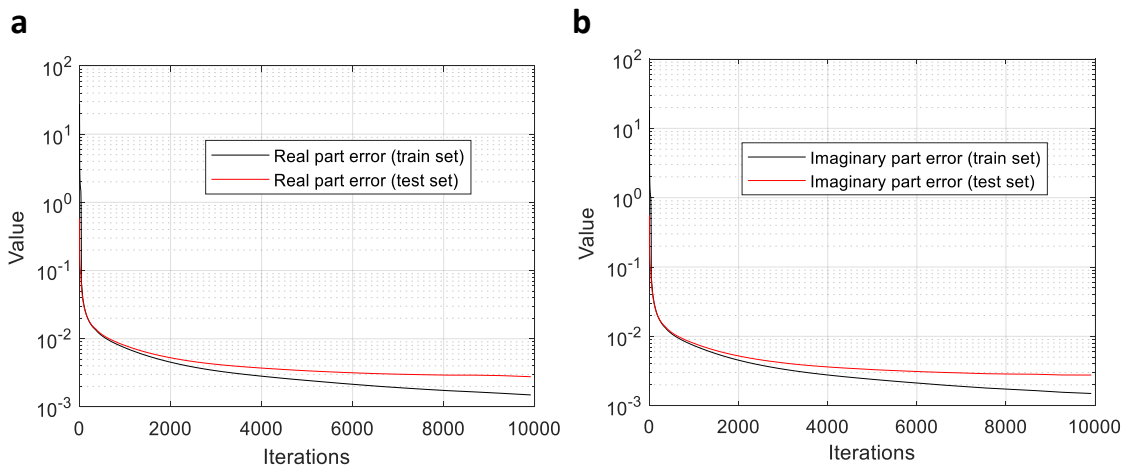

**Figure S3. Learning curves of the PNNs. a** Training error for the PNN with real part component as targets. **b** Training error for the PNN with imaginary part component as targets.

The error history during the training processes are recorded and shown in Fig. S3. Increasing differences between the training data error (in black) and the test data error (in red) are caused by the unavoidable overfitting. To minimize this effect and accelerate the training process, we applied a batch normalization layer after each Convolution-Pooling layer. The ReLU function that was applied to each layer also helped to reduce the overfitting effect. After applying these measures, the overfitting effect is minimized, along with the test data error. As shown in Fig. S3a & Fig. S3b, the test error stabilized after 10,000 epochs of training, indicating the network is well-converged and the training process is over.

**Section IV – Ablation analysis**

To justify the necessity of the NTN layers, batch normalization layers, spatial tiling and split real/imaginary component prediction approach adopted in this PNN, an ablation analysis is also carried out (Fig. S4). We constructed 4 different networks, which all have network structure similar to the original one that is adopted in this paper, but with one configuration removed or modified. Results in Fig. S4 indicate removal or modifications made to any of these configurations would lead to slower convergence or higher final error.

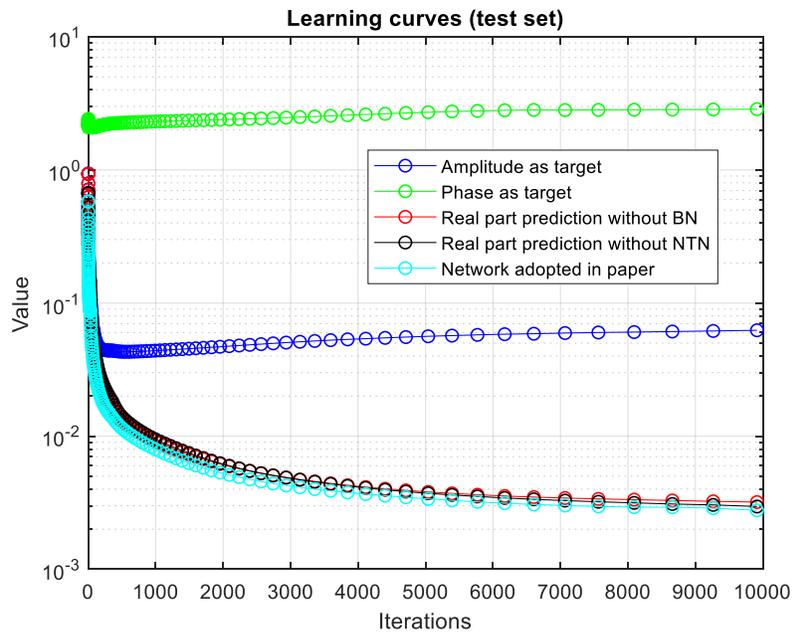

**Figure S4. Ablation analysis.** The mean square error for five different networks are plotted in different colors. All error values are derived with the test dataset. All five networks converged after 10,000 epochs of training. The network structure adopted in this paper has the minimum final error value.

As shown in Fig. S4, for the networks that use amplitude or phase responses as targets, overfitting is apparent from the rising test data error as the training process proceeds. Meanwhile, the two other prediction networks that are constructed without BN layers or NTN layers both converged slower than the PNN adopted in the paper.

## Section V – Additional samples of the PNN

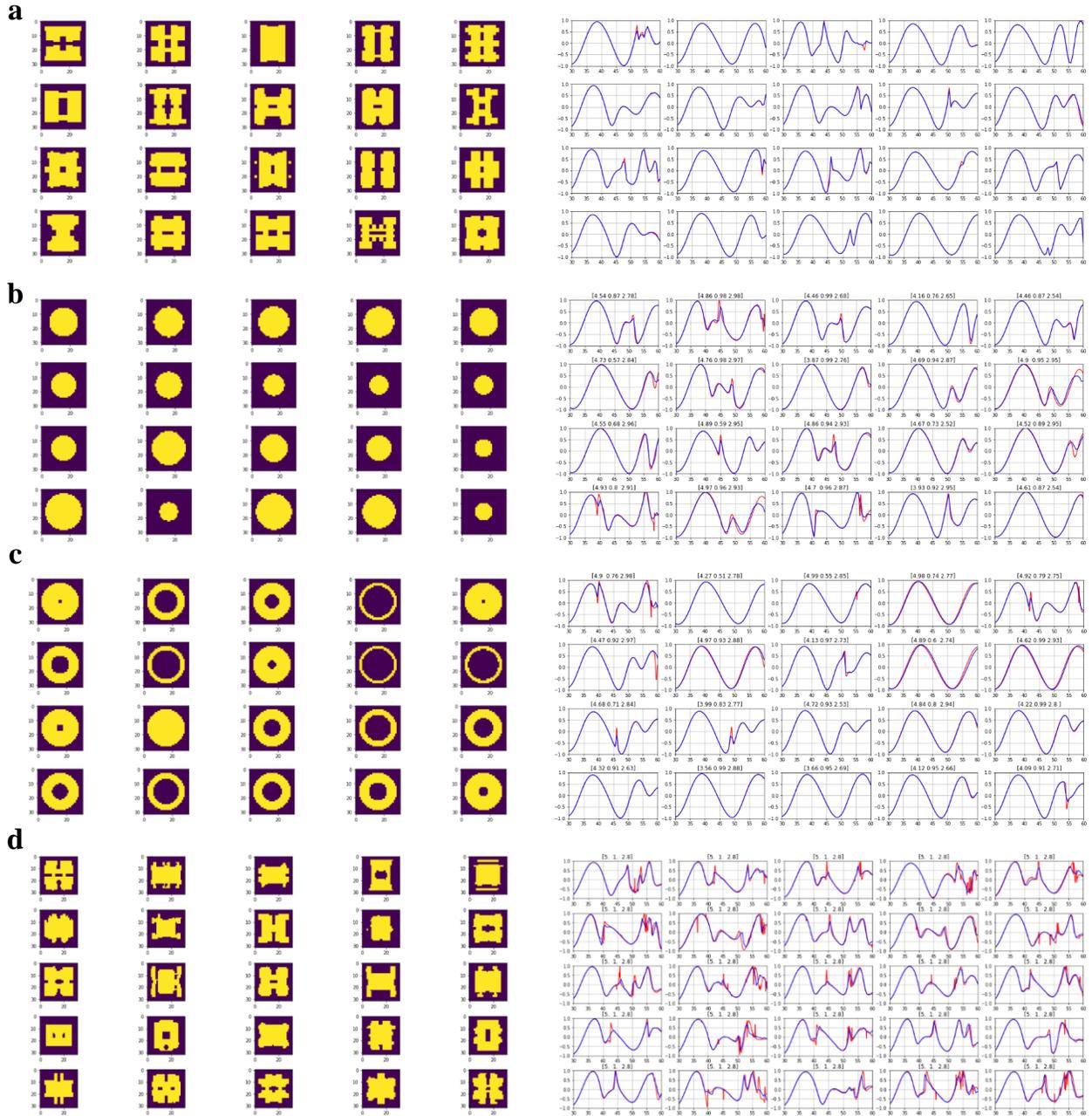

**Figure S5. Additional PNN prediction results compared to accurate results (real parts). a** PNN predictions on meta-atoms selected from the test dataset. **b** PNN predictions on circle-shaped meta-atoms. **c** PNN predictions on ring-shaped meta-atoms. **d** PNN predictions on slightly asymmetric meta-atoms. The refractive index, meta-atom thickness and lattice size, respectively, are shown on top of each subplot (lengths in μm). 2D cross sections of each meta-atom are shown on the left. Only the real parts of the complex transmission coefficients are plotted in this figure. Blue curves represent the PNN predictions, while red curves are simulation results.

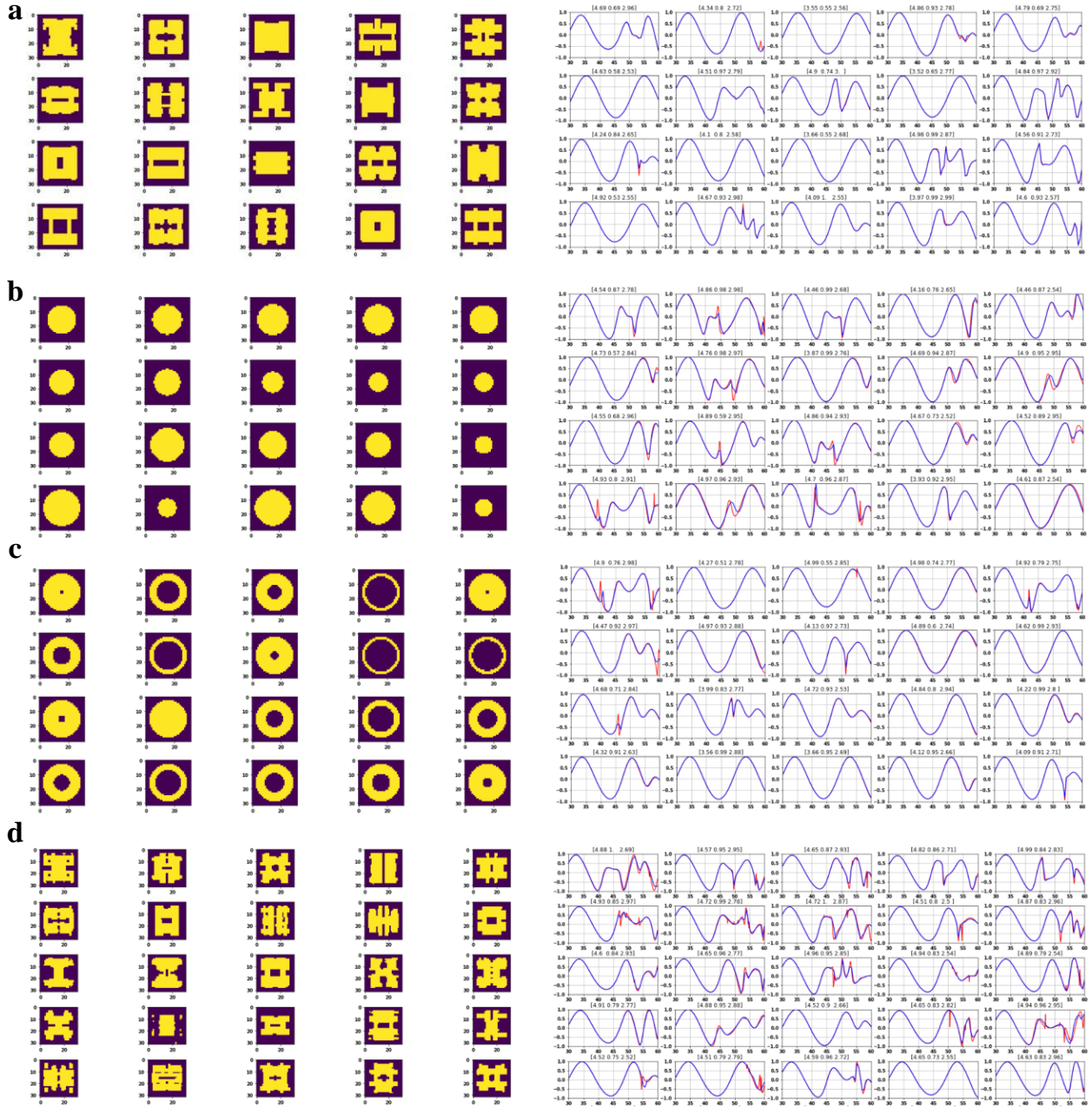

**Figure S6. Additional PNN prediction results compared to accurate results (imaginary parts). a** PNN predictions on meta-atoms selected from test dataset. **b** PNN predictions on circle-shaped meta-atoms. **c** PNN predictions on ring-shaped meta-atoms. **d** PNN predictions on slightly asymmetric meta-atoms. The refractive index, meta-atom thickness and lattice size, respectively, are shown on top of each subplot (lengths in μm). 2D cross sections of each meta-atom are shown on the left. Only the imaginary parts of the complex transmission coefficients are plotted in this figure. Blue curves represent the PNN predictions, while red curves are simulation results.

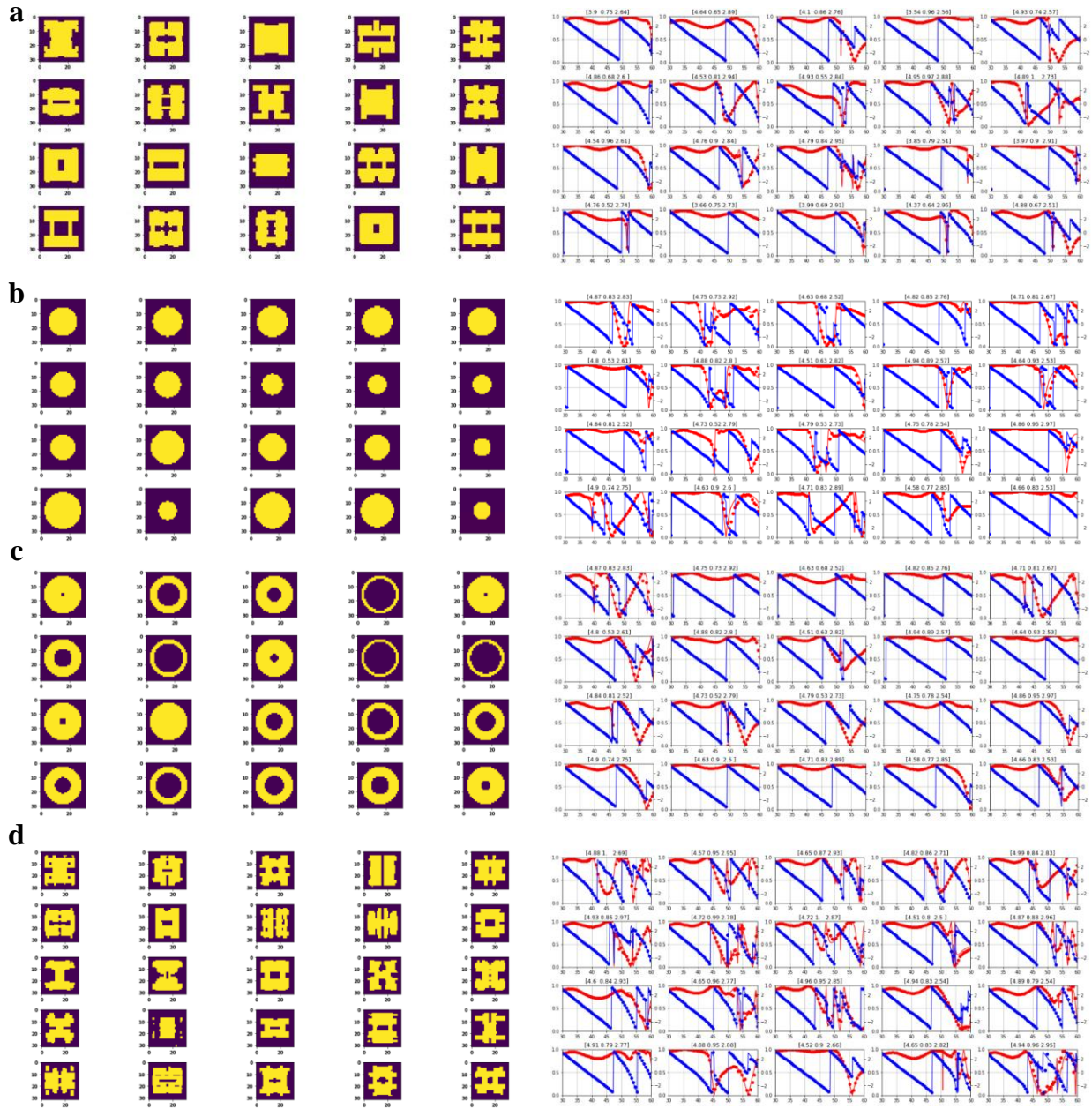

**Figure S7. Additional PNN prediction results compared to accurate results (Amplitude and phase). a** PNN predictions on meta-atoms selected from test dataset. **b** PNN predictions on circle-shaped meta-atoms. **c** PNN predictions on ring-shaped meta-atoms. **d** PNN predictions on slightly asymmetric meta-atoms. The refractive index, meta-atom thickness and lattice size are shown on top of each subplot (lengths in µm). 2D cross sections of each meta-atom are shown on the left. Both amplitude (in red) and phases (in blue) are derived using the real and imaginary part predictions from PNNs. Dotted lines represent the PNN predictions, while solid curves are simulation results.

# Section VI – Details of the metalens optimization process

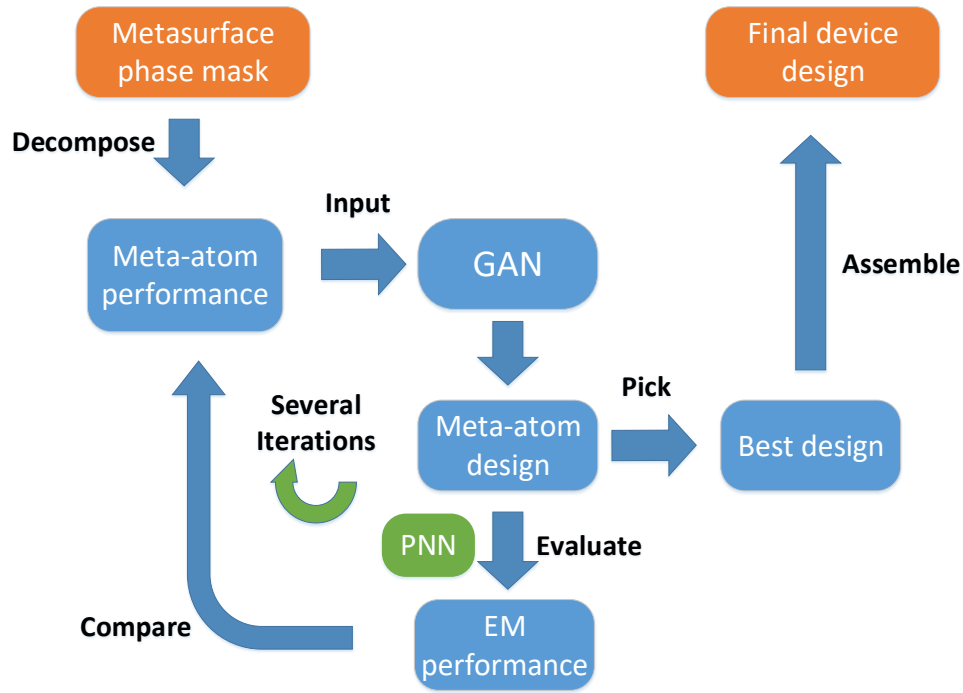

**Figure S8. Flow chart of the lens optimization process.**

Flow chart of the lens optimization process is shown in Fig. S8. Depending on the specific functionalities, the phase mask of the metasurface is calculated and decomposed into distinct meta-atom performance design targets. The well-trained GAN then takes these targets as inputs and generates corresponding meta-atom designs. The EM performance of these generated designs are then evaluated using the well-trained PNN, and the one that has the performance closest to the preset target is chosen to assemble the final device design. The whole optimization process can be consecutively executed for several iterations to maximize the final device's performance.

As mentioned in the paper, we employ this cascaded network (GAN + PNN) to run 1, 2, 5 and 10 times to generate 4 different meta-lens (Fig. S8). The left-top quadrant of each designed metalens is shown in Fig. S9. The whole design can be generated by symmetrically replicating these patterns along the *x* and *y* axes.

## Optimization iteration = 1

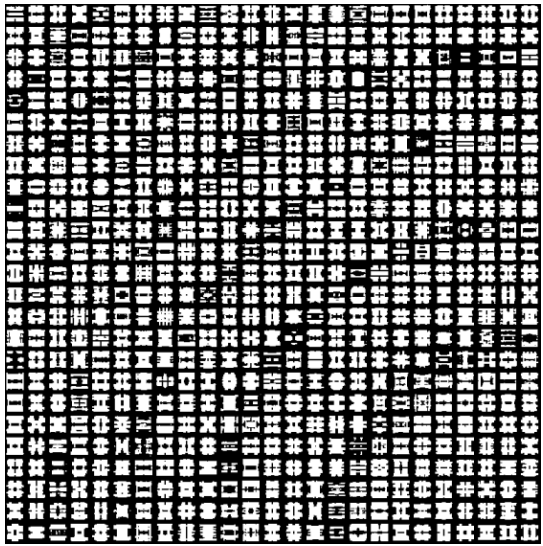

## Optimization iteration = 2

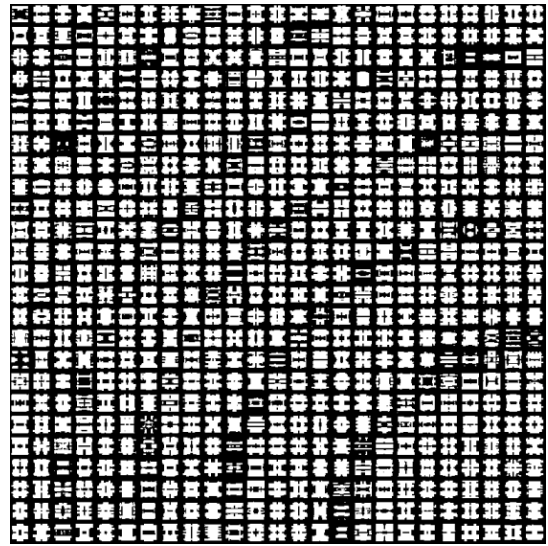

## Optimization iteration = 5

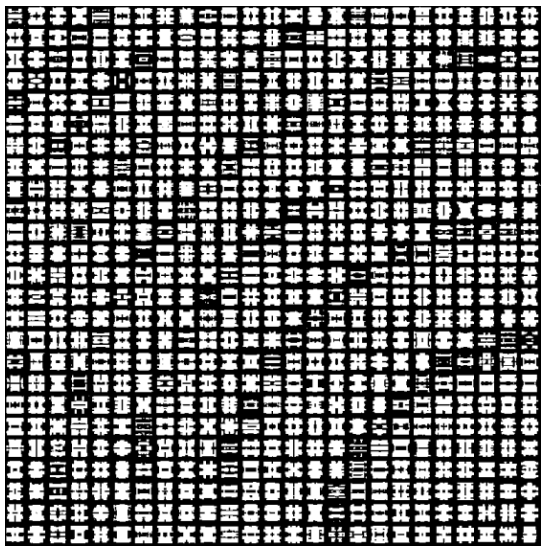

## Optimization iteration = 10

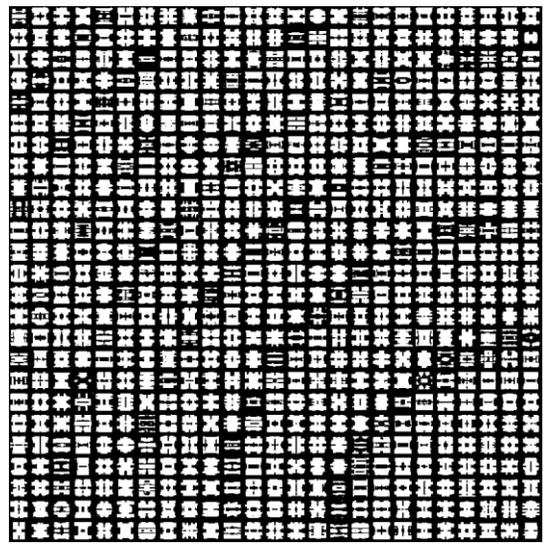

**Figure S9. Four Metalenses optimized using the combined network.**

## Section VII – Computation efficiency analysis

We performed several experiments to compare the efficiency of the PNN with the traditional simulation methods. As shown in Table S2, we measured the time-taken for all validation processes using PNNs in this paper. Then all simulations were performed again using the full-wave simulation tool, CST. From the results showing in Table S2, huge computation time savings are achieved by replacing the conventional method with DNN-based schemes. Importantly, the time taken for the optimization process using PNNs is now comparable with the design time, indicating that the design and optimization of metadevices is able to be achieved on a timescale of seconds using these GAN-PNN cascaded networks. All validations are performed using the same workstation for a fair comparison. The hardware consists of a quad-core CPU with 3.5 GHz clock speed, 64 Gigabytes of RAM and two NVidia 1080Ti GPUs.

Table S2. Time comparisons between DNN-based and traditional methods

| | | With PNNs | With full-wave simulations | Ratio |
|---|---|---|---|---|
| Meta-atom amplitude & phase responses prediction (Single design, 30-60 THz) | | 5.5 ms | 50 s | 9,000 |
| Meta-atom coverage evaluation at one frequency | Fig. 3a (20,000 designs) | 2.15 s | 60 h | 100,000 |
| | Fig. 3b (20,000 designs) | 2.23 s | 60 h | 97,000 |
| Lens optimization (Fig. 4) | Iteration = 1 (625 designs, designed in *0.99* s) | 2.28 s | 2 h | 3,000 |
| | Iteration = 2 (1,300 designs, designed in *1.22* s) | 2.5 s | 4 h | 57,00 |
| | Iteration = 5 (3,125 designs, designed in *1.26* s) | 2.31 s | 10 h | 16,000 |
| | Iteration = 10 (6,250 designs, designed in *1.58* s) | 2.75 s | 20 h | 26,000 |